\documentclass[aps,prl,twocolumn,superscriptaddress]{revtex4}
\usepackage[dvips]{graphicx}
\usepackage{amsmath}
\usepackage{amssymb}
\newcommand{\be}{\begin{equation}}
\newcommand{\ee}{\end{equation}}
\newcommand{\ba}{\begin{eqnarray}}
\newcommand{\ea}{\end{eqnarray}}

\begin{document}

\title{Pulsar kicks via spin-1 color superconductivity}

\author{Andreas Schmitt}
\affiliation{Center for Theoretical Physics, Massachusetts Institute 
of Technology, Cambridge, MA 02139, USA}

\author{Igor A. Shovkovy}
\affiliation{
Frankfurt Institute for Advanced Studies, J.W.\ Goethe-Universit\"at,
D-60054 Frankfurt am Main, Germany}

\author{Qun Wang}
\affiliation{
Institut f\"ur Theoretische Physik, J.W.\ Goethe-Universit\"at,
D-60054 Frankfurt am Main, Germany}

\received{17 February 2005}
\published{2 June 2005}

\begin{abstract}
We propose a new neutrino propulsion mechanism for neutron stars which 
can lead to strong velocity kicks, needed to explain the observed bimodal 
velocity distribution of pulsars. The spatial asymmetry in the neutrino 
emission is naturally provided by a stellar core containing spin-1
color-superconducting quark matter in the A phase. The neutrino propulsion
mechanism switches on when the stellar core temperature drops below the 
transition temperature of this phase. 
\end{abstract}



\maketitle

{\em Introduction.}---The 
first pulsar was discovered more than 37 years ago \cite{pulsars}.
Since then, pulsars remain among the most interesting celestial objects in 
our Galaxy. Their observation allows to test experimentally the existence 
of the gravitational radiation as predicted by general relativity 
\cite{Taylor1989}. The recent discovery of a double-pulsar system 
\cite{Lyne} is likely to further strengthen the status of pulsars as 
an astrophysical laboratory for testing general relativity. In this 
Letter, we suggest that the physics of pulsars may also serve as the
ultimate laboratory for testing the theory of strong interactions, 
quantum chromodynamics (QCD), in the regime of large baryon density.

Matter at large baryon density is expected to be deconfined and 
color-superconducting (for reviews see, e.g., Ref.~\cite{reviews}).
Color-superconducting quark matter could exist 
inside neutron stars, whose central densities are highest 
in Nature. Therefore, it is important to study the physical 
implications of such a possibility in detail. At present, this 
is not easy because our current knowledge regarding the ground 
state of neutral, $\beta$-equilibrated dense matter is very 
limited \cite{phase-d,phase-d1}. Many color superconducting 
phases were proposed \cite{2sc,cfl,gapless,spin-1a,spin-1b,spin-1c},
but it is not clear which of them can be realized inside stars.

It has been known for a long time that typical spatial velocities
of pulsars are an order of magnitude larger than those of their 
progenitors \cite{v-crab,over1000}. Taking into account the violent 
conditions at the pulsar birth, this may not be so surprising. Even 
a small asymmetry in the supernova explosion may result in a kick 
velocity of several hundred $\mbox{km~s}^{-1}$ \cite{Shklovsky}. 
However, the bimodal velocity distribution of pulsars \cite{ACC,BFGHT},
which is unlikely to result from a single physical mechanism, might 
be more surprising. If one associates the low-velocity 
component ($\alt 100~\mbox{km~s}^{-1}$) with asymmetric supernova 
explosions, then what is the origin of the high-velocity component 
($\agt 500~\mbox{km~s}^{-1}$)? Several mechanisms were proposed 
\cite{other,Colpi,Bombaci}, but the issue does not seem to be settled.
For short reviews see, for example, Ref.~\cite{Lai}.

In this Letter, we propose a new neutrino propulsion mechanism, 
resulting from a color superconductor in the transverse A phase 
\cite{andreas}. In this phase, quarks of the same flavor form 
Cooper pairs with total spin one. The neutrino emission from 
this phase, dominated by direct Urca-type processes, is not 
symmetric in space. This emission, as we shall see, provides 
a natural mechanism to power strong, e.g., of order 
$1000~\mbox{km~s}^{-1}$, velocity kicks for neutron stars, which
could explain the high-velocity component in the pulsar distribution 
\cite{ACC,BFGHT}. Unlike most other mechanisms, this one turns 
on not immediately after the supernova explosion, but after the 
temperature of the stellar core drops below the critical temperature 
of the A phase. A distinctive prediction of this mechanism is the 
alignment of the kick velocity direction with the rotational axis.

{\em Neutrino emission.}---Let 
us start by outlining the main steps in the derivation of a 
general expression for the neutrino emissivity in spin-1 color
superconducting phases. We use the Kadanoff-Baym formalism
\cite{KB,Sedrakian} to derive the following differential
expression for the emissivity \cite{long}:
\ba
\frac{d \epsilon_\nu}{dp_\nu d\Omega_\nu}
 &=& \frac{G_F^2}{8(2\pi)^6}
\int \! p_{e} dp_{e} \! \int \! d\Omega_{e}
p_{\nu}^2\,
n_B(p_\nu-p_{e}+\mu_{e}) \nonumber \\
 &\times &
n_F(p_{e}-\mu_{e})\,
L_{\lambda\sigma}(P_{e},P_\nu)\,
\mbox{Im}\Pi_{R}^{\lambda\sigma}(\delta P_{e}-P_\nu),
\nonumber \\
\label{emissivity}
\ea
where $G_F$ is the Fermi coupling constant, $\mu_{e}$ is the electron 
chemical potential, and 
$\delta P_{e}^\lambda \equiv P_{e}^\lambda-\delta_0^\lambda \mu_{e}$. 
Here, particle four-momenta are denoted by capital Latin letters, while
the absolute values of the three-momenta are denoted by lowercase letters. 
The metric tensor is $g_{\lambda\sigma} = \mbox{diag}(1,-1,-1,-1)$.
The Bose and the Fermi distribution functions are denoted by 
$n_B(\omega) \equiv [\exp(\omega/T)-1]^{-1}$ and 
$n_F(\omega) \equiv [\exp(\omega/T)+1]^{-1}$, respectively.
The lepton tensor $L_{\lambda\sigma}(P_{e},P_\nu)$ is defined as follows:
\be
L_{\lambda\sigma}(P_{e},P_\nu)=\mbox{Tr}\left[P_{e}^\kappa\gamma_\kappa 
\gamma_\sigma (1-\gamma^5)P_\nu^\rho \gamma_\rho \gamma_\lambda
(1-\gamma^5)\right].
\ee
Finally, the last factor in the integrand on the right hand side 
of Eq.~(\ref{emissivity}) is the imaginary part of the retarded
polarization tensor of the $W$-boson,
\be
\Pi^{\lambda\sigma}(Q)=\frac{T}{2}
\sum_n \int\frac{d^3 \mathbf{k}}{(2\pi)^3} \mbox{Tr} 
\left[\Gamma^\lambda_{-}S(K)\Gamma^\sigma_{+}S(K+Q)\right],
\label{Pi-mu-nu}
\ee
with the trace running over flavor, color, Dirac and Nambu-Gorkov 
indices. The quark propagator $S(K)$ is diagonal in flavor space, 
and its components have the following Nambu-Gorkov structure:
\be
S_f = \left(\begin{array}{ll}
  G_{f}^{+}(K) & \Xi_{f}^{-}(K) \\
\Xi_{f}^{+}(K) &   G_{f}^{-}(K)
\end{array}\right),\quad \mbox{for}\quad f=u,d.
\label{prop}
\ee
We consider the ultrarelativistic limit. The explicit color and Dirac 
structure of $G_{f}^{\pm}$ and $\Xi_{f}^{\pm}$ can be found in 
Ref.~\cite{andreas}. Here, we note only that the poles of the quark
propagators appear at $k_0=k+\mu_f$ (antiquarks) and at
\be
k_0=\epsilon_{k,r,f}\equiv\sqrt{(k-\mu_f)^2+\lambda_{k,r}|\phi_f|^2},
\quad  r=1,2,3,
\label{modes}
\ee
where $\phi_f$ is the gap parameter, and the functions $\lambda_{k,r}$ 
are specified by the choice of the phase. 

The order parameter of the A phase has a special direction in color 
space: quarks of one color do not pair. Also, it has a special direction 
in momentum space, say, the $z$-direction. If $\theta_\mathbf{k}$
denotes the angle between the three-momentum of a quasiparticle and 
the $z$-axis, the three low-energy quasiparticle modes in the 
transverse A phase are defined by
$\lambda_{k,1}=(1+|\cos\theta_\mathbf{k}|)^2$,  
$\lambda_{k,2}=(1-|\cos\theta_\mathbf{k}|)^2$, and
$\lambda_{k,3}=0$. Here, ``transverse'' refers to the fact that 
quarks of opposite chirality form Cooper pairs.   
 
The explicit expressions of the vertices in Eq.~(\ref{Pi-mu-nu}) 
read
\be
\Gamma^\lambda_{\pm} = \left(\begin{array}{cc}
\gamma^\lambda(1-\gamma^5)  \tau_{\pm} & 0 \\
0 & -\gamma^\lambda (1+\gamma^5) \tau_{\mp}
\end{array}\right),
\label{vert}
\ee
where the flavor matrix $\tau_{\pm}\equiv(\tau_1\pm i\tau_2)/2$ is 
constructed from Pauli matrices. 

By substituting the quark propagators (\ref{prop}) and the vertices 
(\ref{vert}) into Eq.~(\ref{Pi-mu-nu}), we calculate the imaginary
part of the retarded polarization tensor. Then, by making use of the 
result, we derive an expression for the emissivity in the following 
approximate form \cite{long}:
\ba
\epsilon_{\rm tot} &\approx& 2\left( \epsilon^{(11)}_\nu +
\epsilon^{(22)}_\nu +
\epsilon^{(33)}_\nu \right) \nonumber \\
&\approx & \frac{457}{630}
\left[\frac{2}{3}G\left(\frac{\phi_u}{T},\frac{\phi_d}{T}\right)
+\frac{1}{3}\right] \alpha_{s} G_F^2  \mu_e \mu_{u} \mu_{d} T^6,
\label{emis}
\ea
where $\epsilon^{(rr)}_\nu$ is the partial contribution involving 
the $r$th type quasiparticle modes, see Eq.~(\ref{modes}). The 
factor $2$ comes from taking into account both the neutrino and 
the anti-neutrino emissivities. In the final result, the contribution 
of the ungapped modes $\epsilon^{(33)}_\nu$ is, up to a factor $1/3$, 
the same as in the normal phase of quark matter \cite{Iwamoto,Schaf2004}. 
The contribution of the other two modes is suppressed by the 
following function:
\be
G\left(\varphi_u,\varphi_d\right)
\approx \frac{2520}{457\pi^6}
\int_0^\infty d v v^3 \int_{-1}^1 d \xi f(v,\xi),
\ee
where 
\be
f(v,\xi) = \sum_{e_1,e_2=\pm}
\int\limits_{0}^\infty
\int\limits_{0}^\infty 
\frac{\left(
     e^{v+e_{1} \tilde{\epsilon}_{u}
         -e_{2} \tilde{\epsilon}_{d}} +1 \right)^{-1} d x_u d x_d }{
\left(e^{-e_{1} \tilde{\epsilon}_{u}} +1 \right)
\left(e^{ e_{2} \tilde{\epsilon}_{d}} +1 \right)},
\ee
and $\tilde{\epsilon}_{f}=\sqrt{x_f^2+(1+ \xi)^2 \varphi_f^2}$ with $f=u,d$.
Note that $0\leq G\left(\varphi_u,\varphi_d\right)\leq 1$ and $G(0,0)=1$.

In passing, we note that the pair breaking processes \cite{PBP}
do not play any significant role in the case of spin-1 
color-superconducting quark matter under consideration. The 
corresponding contribution to the emissivity \cite{PBP-quark} 
is parametrically suppressed by factor $T/\mu_e\sim 10^{-3}$.

{From} Eq.~(\ref{emis}), we see that the (anti-)neutrino emissivity in 
the A phase differs only by a factor of order 1 from the corresponding 
result in the normal phase of quark matter \cite{Iwamoto,Schaf2004}. Therefore, 
the A phase should have qualitatively the same effect
on cooling of stars as the normal phase.

Nevertheless, the neutrino emission from the A phase is very unusual.
It is not symmetric with respect to reversing the direction of the 
$z$-axis. To quantify the asymmetry, we calculate the value 
of the $z$-component of the momentum carried away by neutrinos per 
unit volume of quark matter, per unit time. This is obtained by 
replacing one power of $p_\nu$ on the right hand side of 
Eq.~(\ref{emissivity}) by $p_\nu \cos\theta_{\mathbf{p}_\nu}$. 
Taking into account that the neutrino and the antineutrino emissions 
give the same contributions, we arrive at the final result \cite{long}
\be
\frac{d P^{\rm (tot)}_z}{dV dt} \approx 
\frac{2}{3}H\left(\frac{\phi_u}{T},\frac{\phi_d}{T}\right)
\frac{457}{630} \alpha_{s} G_F^2  \mu_e \mu_{u} \mu_{d} T^6,
\label{momentum}
\ee
where 
\be
H\left(\varphi_u,\varphi_d\right)
\approx -\frac{840}{457\pi^6}
\int_0^\infty d v v^3 \int_{-1}^1 d \xi \xi f(v,\xi).
\label{H-function}
\ee
Note that $H(0,0)=0$ which is consistent with the fact that the 
momentum kick is vanishing in the normal phase of quark matter. 
The numerical result for the function $H$ at equal values of its 
two arguments is well approximated by the following expression:
\be
H(\varphi,\varphi) \approx  \sum_{n=1}^{5}
\frac{h_n}{\left[1+(r_{0}\varphi)^2\right]^{n/2}},
\label{H-approx}
\ee
with $h_1= 0.3068$,
     $h_2=-0.1977$, 
     $h_3=-0.7838$, 
     $h_4= 1.0286$,
     $h_5=-0.3539$, 
and  $r_0 = e^{\gamma+\bar{\zeta}}/\pi\approx 0.8125$ is the ratio of 
the critical temperature to the value of the gap at $T=0$ in the A phase, 
i.e., $r_0 = T_c/\phi_0$ which is expressed in terms of the Euler 
constant $\gamma\approx 0.577$ and $\bar{\zeta}=\ln2-1/3$ \cite{andreas}. 
Note that $\sum_{n=1}^{5} h_n=0$. The representation in Eq.~(\ref{H-approx})
is particularly convenient when the following simplified temperature 
dependence for the gap parameter is used: 
$\phi(T)=\phi_0\sqrt{1-(T/T_c)^2}$.
In this case, the function $H(T)$ becomes a polynomial: 
    $H(T)=\sum_{n=1}^{5} h_n (T/T_c)^n$ for $T<T_c$, 
and $H(T)=0$ for $T>T_c$.

\begin{figure}[b]
\includegraphics[width=0.4\textwidth]{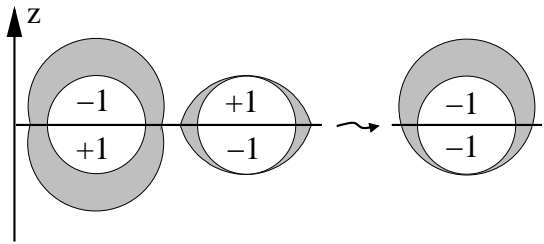}
\caption{\label{effgap}
Gap functions for the first (left) and the second (middle) excitation 
branch with specified helicities of quasiparticles in the upper and 
the lower half-spaces. The ``effective'' gap relevant for the neutrino 
emission is shown on the right.}
\end{figure}

The physical reason for the breakdown of the reflection symmetry 
lies in the pairing pattern of the transverse A phase. Quasiquarks 
of the first branch, $r=1$, have helicity $+1$ if the projection 
of their momentum onto the $z$-axis is negative, $\cos\theta_{\bf k}<0$, 
and helicity $-1$ if $\cos\theta_{\bf k}>0$. Quasiquarks of the second 
branch, $r=2$, have opposite helicities. Only left-handed quarks (in 
the ultrarelativistic limit, quarks with negative helicity) participate 
in the Urca processes. Thus, the quasiquarks of the first (second) 
branch contribute only if their momenta are in the upper (lower) 
half-space. Taking this into account, the effective branch relevant 
for the emission has gap $\phi_{\rm eff}\sim 1 + \cos\theta_{\bf k}$, 
which discriminates between $+z$ and $-z$ directions, see 
Fig.~\ref{effgap}. Since neutrinos are emitted preferably in the 
direction opposite to the quark momenta, this asymmetry manifests 
itself in the neutrino emission.     

{\em Calculation of the velocity kick.}---In 
order to make a simple estimate of the velocity kick due to
the neutrino emission from a  color-superconducting quark matter
core in the A phase, we use the following model time dependence 
of the core temperature (see, e.g., Ref.~\cite{Schaf2004}):
$T(t)=T_0(t_0/t)^{1/4}$. This has the power dependence which is 
characteristic for bulk matter cooling by neutrino emission with 
$\epsilon_\nu \sim T^6$, see Eq.~(\ref{emis}), provided the 
specific heat of matter is $c_V\sim T$. Here, we assume that 
$T_c<T_0$, i.e., the system is too hot for spin-1 pairing 
initially, and then it cools through the transition point.

The velocity kick for a star of mass $1.4 M_\odot$ with the 
quark core of radius $R_c$ is given by the expression:
\ba
\delta v 
&\equiv& \frac{\Delta P_{z}^{\rm (tot)}}{1.4 M_\odot}
=\frac{457\alpha_s}{945} G_F^2
\mu_{e} \mu_{u} \mu_{d} 
\frac{4\pi}{3} \frac{R_c^3}{1.4 M_\odot} T_0^4 T_c^2 t_0
\nonumber\\
&\times& \theta(t -t_c) \sum_{n=1}^{5}\frac{4h_n}{2+n}
\left[1-\left(\frac{t_c}{t }\right)^{(2+n)/4}\right],
\label{kick-vel}
\ea
where we used the approximate expression in Eq.~(\ref{H-approx}),
as well as a simplified temperature dependence of the gap
parameters, $\phi_u(T)=\phi_d(T)=\phi_0\sqrt{1-(T/T_c)^2}$. The 
notation $t_c\equiv t_0 (T_0/T_c)^4$ stands for the time when 
the temperature of the quark core drops below the critical 
value. 

\begin{figure}[b]
\includegraphics[width=0.45\textwidth]{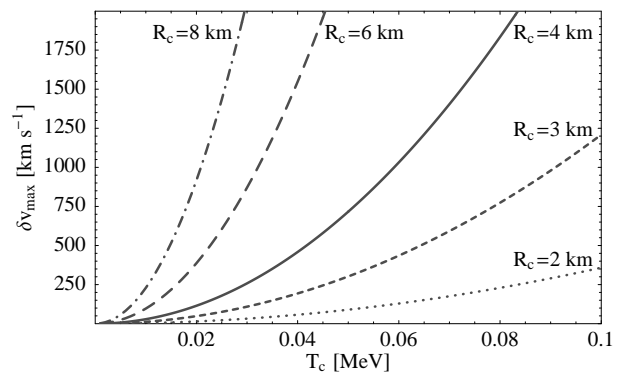}
\caption{\label{velocity-vs-Tc}
The dependence of the maximum velocity kick of a neutron star 
versus the value of the critical temperature of the spin-1 color
superconducting phase transition. Results for several values of 
the quark core radius are shown.}
\end{figure}

{From} Eq.~(\ref{kick-vel}), we can also derive the expression
for the maximum velocity kick ($t =\infty$):
\be
\delta v_{\rm max} \approx 0.033 \,\alpha_s G_F^2
\mu_{e} \mu_{u} \mu_{d} 
\frac{4\pi}{3} \frac{R_c^3}{1.4 M_\odot} T_0^4 T_c^2 t_0.
\ee 
The results for the maximum velocity kicks are presented graphically 
in Fig.~\ref{velocity-vs-Tc}. To make the plot, we take $\alpha_s = 1$,
and use the initial condition: 
$T_0 = 100 \mbox{~keV}$
at 
$t_0=100 \mbox{~yr}$. 
Also, we use the following values of the chemical potentials:
$\mu_{u} = 400 \mbox{~MeV}$,
$\mu_{d} = 500 \mbox{~MeV}$,
$\mu_{e} = 100 \mbox{~MeV}$.

When the stellar age $t $ is finite, all curves in 
Fig.~\ref{velocity-vs-Tc} shift to the right. Numerically,
the value of the shift is $10\mbox{~keV}$ at $t =10^6\mbox{~yr}$ 
and about $\mbox{30~keV}$ at $t =10^4\mbox{~yr}$. This is easy 
to understand: at $t <\infty$, the step function in 
Eq.~(\ref{kick-vel}) cuts out a range of low values of critical 
temperatures, $T_c<T_0(t_0/t )^{1/4}$, which are not yet accessible 
at time $t $. Other than this shift, the shape of the curves remains 
similar even at $t $ as low as $10^3\mbox{~yr}$. 

{\em Discussion.}---The 
main observation of this Letter is that the neutrino emission 
from the spin-1 color-superconducting transverse A phase (unlike
from other spin-1 color-superconducting phases \cite{long}) is
not symmetric with respect to reversing one spatial direction. 

Therefore, we propose a hypothesis according to which the inner cores 
of some (hybrid) neutron stars are made of the A phase. The neutrino 
emission from such stars could generate strong velocity kicks, needed
to explain the observed bimodal velocity distribution of pulsars 
\cite{ACC,BFGHT}. Its bimodal structure results from an overlap 
of two distributions: one describing ``normal'' neutron stars, and 
the other describing hybrid stars with color-superconducting quark 
cores in the A phase.

The above hypothesis has several specific (and, thus, falsifiable)
predictions, which are directly related to the nature of the  
A phase. Let us start by discussing the 
direction of the kick velocity. As should be clear from our analysis,
this is aligned with the direction spontaneously picked by the spin-1
condensate. In the case of an infinite isotropic medium, this 
direction can be arbitrary. Inside a rotating and magnetized star, 
however, the degeneracy with respect to the orientation is likely 
to be removed because of a nonvanishing interaction with the angular 
momentum and/or with the magnetic field. Therefore, the kick velocity 
averaged over time should be parallel to the rotational axis. The 
observational data seems to favor this possibility \cite{Romani}.

The proposed neutrino propulsion mechanism also predicts a
correlation between the velocity distribution and the age 
of young neutron stars. This has not been seen in the 
observational data yet \cite{ACC}. Most likely, however, this 
correlation is hard to detect because it takes typically less
than $10^4~\mbox{yr}$ for a star to accumulate almost all of 
the maximum velocity kick. The future high statistics studies 
of young neutron stars may resolve the issue.

Along with the advantages of the proposed mechanism, it might 
be also appropriate to mention several potential difficulties.
One of them is a rigorous justification that the spin-1
color-superconducting A phase is the ground state of dense quark 
matter inside a star. In the case of an infinite isotropic medium, 
it has been shown that the color-spin locked phase is the favored 
spin-1 phase at asymptotically large density \cite{spin-1b,andreas}. 
We could only speculate that this may change when the density is 
realistic, and the rotation and the magnetic field are accounted for. 
The other potential difficulty is related to fast cooling of the 
A phase by direct Urca processes, which might not be compatible
with the observational data \cite{cooling}.

The mechanism, proposed here, may well have other important
implications that we did not discuss in this Letter. We hope, 
however, that they will be addressed in the future studies, 
providing much stronger tests for the hypothesis of the hybrid 
stars accelerated by the asymmetric neutrino emission from the
A phase. If it passes the tests, the observational data from 
pulsars could shed light on some details of the QCD phase 
diagram. Thus, the pulsars should be viewed not only as a unique
laboratory for testing the theory of general relativity, but 
also the ultimate laboratory for testing the theory of QCD.

\begin{acknowledgments}
The authors thank D. Rischke for suggesting this topic. They also thank
D. Blaschke, S. Popov, S. Reddy, A. Sedrakian, and J. Schaffner-Bielich
for stimulating discussions on related topics. This work was supported
in part by the Virtual Institute of the Helmholtz Association under
grant No. VH-VI-041 and by Gesellschaft f\"{u}r Schwerionenforschung
(GSI), Bundesministerium f\"{u}r Bildung und Forschung (BMBF).
A.S. thanks the German Academic Exchange Service (DAAD) for
financial support and the MIT Center for Theoretical Physics 
for its kind hospitality.
\end{acknowledgments}


\begin{thebibliography}{99}

\bibitem{pulsars} A.~Hewish, S.~J.~Bell, J.~D.~H.~Pilkington,  P.~F.~Scott, 
and R.~A.~Collins, 
Nature (London) {\bf 217}, 709 (1968).

\bibitem{Taylor1989}
J.~H.~Taylor and J.~M.~Weisberg,
Astrophys.\ J.\  {\bf 345}, 434 (1989);
J.~M.~Weisberg and J.~H.~Taylor,
astro-ph/0407149.

\bibitem{Lyne}
A.~G.~Lyne {\it et al.},
Science {\bf 303}, 1153 (2004). 

\bibitem{reviews}  K.~Rajagopal and F.~Wilczek,
hep-ph/0011333;
M.~Alford,
Annu.\ Rev.\ Nucl.\ Part.\ Sci.\  {\bf 51}, 131 (2001);
S.~Reddy,
Acta Phys.\ Pol.\ B {\bf 33}, 4101 (2002).
T.~Sch{\"a}fer,
hep-ph/0304281;
D.~H.~Rischke,
Prog.\ Part.\ Nucl.\ Phys.\  {\bf 52}, 197 (2004);
M.~Buballa,
Phys.\ Rep.\ {\bf 407}, 205 (2005);
H.-C.~Ren,
hep-ph/0404074;
M.~Huang,
hep-ph/0409167;
I.~A.~Shovkovy,
nucl-th/0410091.

\bibitem{phase-d}
S.~B.~R\"{u}ster, I.~A.~Shovkovy and D.~H.~Rischke,
Nucl.\ Phys.\ A {\bf 743}, 127 (2004);
nucl-th/0411040.

\bibitem{phase-d1} 
K.~Fukushima, C.~Kouvaris and K.~Rajagopal,
Phys.\ Rev.\ D {\bf 71}, 034002 (2005).

\bibitem{2sc}
M.~Alford, K.~Rajagopal, and F.~Wilczek,
Phys.\ Lett.\ B {\bf 422}, 247 (1998);
R.~Rapp, T.~Sch{\"a}fer, E.~V.~Shuryak, and M.~Velkovsky,
Phys.\ Rev.\ Lett.\  {\bf 81}, 53 (1998).

\bibitem{cfl} 
M.~G.~Alford, K.~Rajagopal, and F.~Wilczek,
Nucl.\ Phys.\ {\bf B537}, 443 (1999).

\bibitem{gapless}
I.~Shovkovy and M.~Huang,
Phys.\ Lett.\ B {\bf 564}, 205 (2003);
M.~Huang and I.~Shovkovy,
Nucl.\ Phys.\ A {\bf 729}, 835 (2003);
M.~Alford, C.~Kouvaris, and K.~Rajagopal,
Phys.\ Rev.\ Lett.\  {\bf 92}, 222001 (2004);
hep-ph/0406137.

\bibitem{spin-1a}
M.~Iwasaki and T.~Iwado,
{\it Phys. Lett. B} {\bf 350}, 163 (1995).

\bibitem{spin-1b}
T.~Sch{\"a}fer,
Phys.\ Rev.\ D {\bf 62}, 094007 (2000).

\bibitem{spin-1c}
M.~Buballa, J.~Ho\v{s}ek and M.~Oertel,
Phys.\ Rev.\ Lett.\  {\bf 90}, 182002 (2003);
A.~Schmitt, Q.~Wang and D.~H.~Rischke,
Phys.\ Rev.\ D {\bf 66}, 114010 (2002);
Phys.\ Rev.\ Lett.\  {\bf 91}, 242301 (2003);
A.~Schmitt,
nucl-th/0405076.

\bibitem{v-crab} V.~Trimble,
Astron. J. {\bf 73}, 535 (1968).

\bibitem{over1000} A.~G.~Lyne and D.~R.~Lorimer,
Nature (London) {\bf 369}, 127 (1994).

\bibitem{Shklovsky} I.~S.~Shklovsky,
Sov. Astron. {\bf 13}, 562 (1970).

\bibitem{ACC}
Z.~Arzoumanian, D.~F.~Chernoff, and J.~M.~Cordes,
Astrophys.\ J.\  {\bf 568}, 289 (2002);
J.~M.~Cordes and D.~F.~Chernoff,
Astrophys.\ J.\  {\bf 505}, 315 (1998).

\bibitem{BFGHT} W.~F.~Brisken, A.~S.~Fruchter, W.~M.~Goss,
R.~S.~Herrnstein and S.~E.~Thorsett,
Astron.\ J.\ {\bf 126}, 3090 (2003).

\bibitem{other}
A.~Kusenko and G.~Segr\`{e},
Phys.\ Rev.\ Lett.\  {\bf 77}, 4872 (1996).

\bibitem{Colpi} M.~Colpi and I.~Wasserman,
Astrophys.\ J.\  {\bf 581}, 1271 (2002).

\bibitem{Bombaci}
I.~Bombaci and S.~B.~Popov,
Astron.\ Astrophys.\  {\bf 424}, 627 (2004).

\bibitem{Lai}
D.~Lai,
astro-ph/0312542;
A.~Kusenko,
Int.\ J.\ Mod.\ Phys.\ D {\bf 13}, 2065 (2004);
C.~J.~Horowitz,
nucl-th/0410074.

\bibitem{andreas} A.~Schmitt,
Phys.\ Rev.\ D {\bf 71}, 054016 (2005).

\bibitem{KB} L.~P.~Kadanoff and G.~Baym,
{\it Quantum Statistical Mechanics}
(Benjamin, New York, 1962).

\bibitem{Sedrakian}
A.~Sedrakian and A.~Dieperink,
Phys.\ Lett.\ B {\bf 463}, 145 (1999);
Phys.\ Rev.\ D {\bf 62}, 083002 (2000).

\bibitem{long} A.~Schmitt, I.~A.~Shovkovy, and Q.~Wang,
(to be published).

\bibitem{Iwamoto}
N.~Iwamoto,
Phys.\ Rev.\ Lett.\  {\bf 44}, 1637 (1980).

\bibitem{Schaf2004}
T.~Sch\"afer and K.~Schwenzer,
Phys.\ Rev.\ D {\bf 70}, 114037 (2004).

\bibitem{PBP} E. Flowers, M. Ruderman, and P. Sutherland,
Astrophys. J. {\bf 205}, 541 (1976);
D.~N.~Voskresensky and A.~V.~Senatorov,
Sov.\ Phys.\ JETP {\bf 63}, 885 (1986).

\bibitem{PBP-quark} 
P.~Jaikumar and M.~Prakash,
Phys.\ Lett.\ B {\bf 516}, 345 (2001).

\bibitem{Romani}
R.~W.~Romani,
astro-ph/0404100.

\bibitem{cooling}
H.~Grigorian, D.~Blaschke and D.~Voskresensky,
Phys.\ Rev.\ C {\bf 71}, 045801 (2005).

\end{thebibliography}
\end{document}


\title{Erratum: Pulsar kicks via spin-1 color superconductivity\\
{[Phys. Rev. Lett. 94, 211101 (2005)]}}

\author{Andreas Schmitt}

\author{Igor A. Shovkovy}

\author{Qun Wang}

\received{15 August 2005}
\published{6 October 2005}
\maketitle

The model time dependence of the temperature, 
\be
T(t)=T_0\left(\frac{t_0}{t}\right)^{1/4},
\ee
which was used in our calculation, is unphysical. As a 
result, the estimate of the kick velocities due to spin-one 
color superconductivity was wrong. 

The correct time dependence is approximately given by
\be
T(t)=T_0\frac{\tau^{1/4}}{\left(t-t_0+\tau\right)^{1/4}},
\ee
which is the same as for normal quark matter. This time 
dependence is obtained by integrating the energy balance 
equation, $\epsilon_{\nu}=-c_{V} dT/dt$, in which an approximate
analytical expression for the specific heat of quark matter,
$c_V=(\mu_u^2+\mu_d^2)T$ (see, for example, Ref.~\cite{Schafer}) 
and Iwamoto's analytical expression for the neutrino emissivity 
$\epsilon_{\nu}$ \cite{Iwamoto} were used. This is justified for 
the long-term stellar cooling after the thermal relaxation 
epoch is completed ($t\agt 10^2$~yr) and before the surface 
cooling by photons starts to dominate ($t\alt 10^5$~yr), e.g., 
see Ref.~\cite{Yakovlev}. In the above expression, $\tau$ is 
defined by  
\be
\tau\equiv
\frac{315}{914} \,\frac{\mu_u^2 + \mu_d^2}
{\alpha_s\, G_F^2 \,\mu_e \,\mu_u \,\mu_d}\,\frac{1}{T_0^4} 
\approx 10^{-5}~\mbox{yr}. 
\ee
Taking this into account, Eqs.~(13) and (14) in our Letter should 
be replaced by 
\be
\delta v 
\equiv \frac{\Delta P_{z}^{\rm (tot)}}{1.4 M_\odot}
=\frac{457\alpha_s}{945} G_F^2
\mu_{e} \mu_{u} \mu_{d} 
\frac{4\pi}{3} \frac{R_c^3}{1.4 M_\odot} T_0^4\, T_c^2 \, \tau\,
\theta(t -t_c) \sum_{n=1}^{5}\frac{4h_n}{2+n}
\left[1-\left(\frac{t_c-t_0+\tau}{t-t_0+\tau}\right)^{(2+n)/4}\right],
\label{kick-vel}
\ee
and 
\be
\delta v_{\rm max} \approx 0.033 \,\alpha_s G_F^2
\mu_{e} \mu_{u} \mu_{d} 
\frac{4\pi}{3} \frac{R_c^3}{1.4 M_\odot} T_0^4 \,T_c^2 \,\tau,
\ee 
respectively. Here $t_c \equiv \tau (T_0^4/T_c^4 -1)+t_0$ is the time when the
temperature is equal to the critical value.

The revised expression for $\delta v_{\rm max}$ differs by a factor
$\tau/t_0\approx 10^{-7}$ from the old one. Consequently, the predicted kick 
velocities become extremely small. For example, instead of velocities 
of the order of $\delta v_{\rm max} \sim 1000$ km/s, one gets
$\delta v_{\rm max} \sim 10^{-4}$ km/s.